\documentclass[a4paper]{article}
\usepackage{geometry}
\usepackage{setspace}
\usepackage{float}
\usepackage{epstopdf}
\usepackage{standalone}
\usepackage{amsmath,amsfonts,amssymb}
\newtheorem{assumption}{Assumption}     
\newtheorem{theorem}{Theorem}    
 
\usepackage{bm}
\usepackage{pdflscape}
\usepackage{authblk}
\usepackage{graphicx}
\usepackage[flushleft]{threeparttable}
\usepackage[comma,authoryear]{natbib}

\usepackage{mathtools}

\newcommand{\interior}[1]{%
 {\kern0pt#1}^{\mathrm{o}}%
}
\usepackage [english]{babel}
\usepackage [autostyle, english = american]{csquotes}
\MakeOuterQuote{"}

\usepackage{algorithm,algpseudocode}
\usepackage{caption}
\usepackage[makeroom]{cancel}

\title{\textbf{Bayesian Indirect Inference for Models with Intractable Normalizing Functions}}
\author{Jaewoo Park}
\affil{Department of Applied Statistics, Yonsei University}

\begin{document}

\maketitle

\begin{abstract}
Inference for doubly intractable distributions is challenging because the intractable normalizing functions of these models include parameters of interest. Previous auxiliary variable MCMC algorithms are infeasible for multi-dimensional models with large data sets because they depend on expensive auxiliary variable simulation at each iteration. We develop a fast Bayesian indirect algorithm by replacing an expensive auxiliary variable simulation from a probability model with a computationally cheap simulation from a surrogate model. We learn the relationship between the surrogate model parameters and the probability model parameters using Gaussian process approximations. We apply our methods to challenging simulated and real data examples, and illustrate that the algorithm addresses both computational and inferential challenges for doubly intractable distributions. Especially for a large social network model with 10 parameters, we show that our method can reduce computing time from about 2 weeks to 5 hours, compared to the previous method. Our method allows practitioners to carry out Bayesian inference for more complex models with larger data sets than before. 
\end{abstract}

\noindent%


{\it Word count: 4683 words }
\vfill

\clearpage
\section{Introduction}

~~~~Models with intractable normalizing functions are common in social science, ecology, epidemiology and physics, among other disciplines. Examples include exponential random graph models \citep[]{robins2007introduction} for social networks, autonormal models \citep[]{hughes2011autologistic} in spatial statisitcs, and interaction spatial point process models \citep[]{goldstein2014attraction}. Consider $h(\mathbf{x}|\bm{\theta})$, an unnormalized likelihood function for a random variable $\mathbf{x} \in \mathcal{X}$ given a parameter vector $\bm{\theta} \in \bm{\Theta}$. In many cases, $h(\mathbf{x}|\bm{\theta})$ takes the form $\exp(\bm{\theta}^\intercal S_\mathbf{x})$, where $S_\mathbf{x}$ is the vector valued sufficient statistics. Let $p(\bm{\theta})$ be the prior density and $\exp(\bm{\theta}^\intercal S_\mathbf{x})/Z(\bm{\theta})$ be the likelihood function. Then, the posterior density of $\bm{\theta}$ is

\begin{equation}
\pi(\bm{\theta}|\mathbf{x}) \propto p(\bm{\theta})\frac{\exp(\bm{\theta}^\intercal S_\mathbf{x})}{Z(\bm{\theta})}.
\label{e1}
\end{equation}

\noindent This results in posterior distributions with extra unknown normalizing terms than in standard Bayesian analysis. This is called as "doubly intractable distributions." The major issue for these models is that $Z(\bm{\theta})$ cannot be easily evaluated. Several algorithms have been developed to address such challenges either by plugging Monte Carlo approximation of $Z(\bm{\theta})$ \citep[cf.][]{atchade2008bayesian, lyne2015russian, park2020function} or by simulating auxiliary variable to avoid direct evaluation of $Z(\bm{\theta})$ \citep[cf.][]{moller2006efficient, murray2006, liang2010double}. The recent Bayes approaches in this field is reviewed by \cite{park2018bayesian}. However, for large data sets, both approaches become computationally expensive. This is because all the algorithms require sampling from the probability model either for a Monte Carlo approximation or for an auxiliary variable generation. To address such challenges, we propose a Bayesian indirect inference for such models via fast simulations from a surrogate model. We show how this new approach is fast while producing accurate posterior approximation.

Recently, several efficient precomputation approaches have been proposed for doubly intractable distributions. For instance, \cite{boland2018efficient} develops fast Monte Carlo approximations for $Z(\bm{\theta})$. They construct design points over parameter space via a stochastic approximation algorithm \citep{robbins1951stochastic}. Then a number of auxiliary variables are generated for each design point to construct an importance sampling estimate of $Z(\bm{\theta})$ before implementing an MCMC algorithm. \cite{moores2015pre} develops a novel indirect approach based on preprocessing for approximate Bayesian computation (ABC) methods \citep{beaumont2002approximate}. For a given parameter, \cite{moores2015pre} generates summary statistics from the surrogate normal distribution. The parameter is accepted if this simulated summary statistic is close enough to the observed summary statistic. These approaches can reduce computing time by avoiding expensive simulations from the model with each iteration of the MCMC algorithm. Our method is motivated by these recent scalable precomputing approaches. Our approach has similarities to \cite{moores2015pre} in that we develop Bayesian indirect approach by generating summary statistics from the surrogate normal distribution. \cite{moores2015pre} applies their method for an 1-dimensional model with univariate interpolation function. In contrast, our approach is generally applicable to moderate dimensional parameter space (up to 10) with large data sets. 



Our method is based on the auxiliary variable MCMC algorithm to avoid direct approximation of $Z(\bm{\theta})$, which is demanding for multi-dimensional $\bm{\theta}$. In our model, generating the auxiliary variable is equivalent to generating the sufficient statistics of the auxiliary variable. Here, we approximate the distribution of the sufficient statistics via a surrogate normal distribution. Intuition comes from that most of the sufficient statistics take some form of summation of elements in data $\mathbf{x}$. Though the central limit theorem does not hold rigorously, our normal approximation appears to provide reasonable estimates as we investigate. Our method may be summarized as follows: (1) approximate mean and covariance of the sufficient statistics at several $\bm{\theta}$ values, and (2) interpolate the mean of the sufficient statistics at other parameter values using a Gaussian process fit. These two steps allow constructing surrogate normal distribution. Then, with each iteration of the MCMC algorithm, we simulate the sufficient statistics from the surrogate normal distribution. We show how our method may be useful in addressing computational challenges for doubly intractable posterior distributions with large data sets.

The remainder of this manuscript is organized as follows. In Section 2, we describe auxiliary variable MCMC approaches for doubly-intractable distributions and point out their computational challenges. In Section 3, we propose a computationally efficient indirect auxiliary variable approach and provide implementation details. In Section 4, we study the performance of our approach in the context of three different examples. We study the computational and statistical efficiency of our approach by comparing it with an existing algorithm. We conclude with a summary and discussion in Section 5.


\section{Auxiliary Variable MCMC for Doubly Intractable Distributions}

~~~~Auxiliary variable approaches \citep{moller2006efficient,murray2006} construct a joint target distribution which includes both model parameters and an auxiliary variable. Then, they update the augmented state via the Metropolis-Hastings algorithm. Let $\mathbf{y}$ be the auxiliary variable which follows the probability model $\exp(\bm{\theta_{\ast}}^\intercal S_\mathbf{y})/Z(\bm{\theta_{\ast}})$, where $S_\mathbf{y}$ is the sufficient statistics of $\mathbf{y}$. The conditional density of $\bm{\theta_{\ast}}$ given $\bm{\theta}$ is $q(\bm{\theta}_{\ast}|\bm{\theta})$. Then, the joint target distribution is

\begin{equation}
\pi(\bm{\theta},\bm{\theta_{\ast}},\mathbf{y}|\mathbf{x}) \propto p(\bm{\theta})\frac{\exp(\bm{\theta_{\ast}}^\intercal S_\mathbf{x})}{Z(\bm{\theta})}q(\bm{\theta_{\ast}}|\bm{\theta})\frac{\exp(\bm{\theta_{\ast}}^\intercal S_\mathbf{y})}{Z(\bm{\theta_{\ast}})}.
\label{jointtarget}
\end{equation}

\noindent For this augmented state, $\bm{\theta_{\ast}}$ is generated from $q(\bm{\theta_{\ast}}|\bm{\theta})$ and the auxiliary variable $\mathbf{y}$ is generated from the probability model  $\exp(\bm{\theta_{\ast}}^\intercal S_\mathbf{y})/Z(\bm{\theta_{\ast}})$. Then, the intractable normalizing functions get canceled out in the acceptance probability of the Metropolis-Hastings algorithm. Several auxiliary variable MCMC (AVM) algorithms have been developed and the difference between these algorithms is how the auxiliary variable is generated. 

The exchange algorithm \citep{moller2006efficient,murray2006} generates an auxiliary variable from the probability model using a perfect sampling \citep{propp1996exact}, which uses bounding Markov chains to generate a random variable exactly. These algorithms are asymptotically exact in that the stationary distribution of the Markov chain is equal to the desired target posterior distribution. However, perfect samplers are only available for some models (e.g., Markov random field model with small size data). This is a major practical issue for their approach. To address such challenges, \cite{liang2010double} develops the double Metropolis-Hastings (DMH) algorithm by replacing a perfect sampler with a standard MCMC sampler. With each iteration of the MCMC update, an auxiliary variable is generated via another MCMC algorithm. DMH is asymptotically inexact because auxiliary variables are generated approximately from a standard MCMC sampler. However, among current approaches, it is the most practical approach in terms of the effective sample size per time \citep{park2018bayesian}. Since generating $\mathbf{y}$ is equivalent to generating $S_\mathbf{y}$, the DMH algorithm may be written as Algorithm~\ref{DMHalg}.

\begin{algorithm}
\caption{Double Metropolis-Hastings algorithm}\label{DMHalg}
\begin{algorithmic}
\normalsize
\State Given $\bm{\theta}_{n} \in \bm{\Theta}$ at $n$th iteration.\\

\State {\it{Step 1.}} Propose $\bm{\theta}_{\ast} \sim q(\cdot|\bm{\theta}_{n})$.\\

\State {\it{Step 2.}} Generate the sufficient statistics of an auxiliary variable approximately from probability model at $\bm{\theta}_{\ast}$: $S_\mathbf{y} \sim \exp(\bm{\theta_{\ast}}^\intercal S_{\mathbf{y}})$ using Metropolis-Hastings updates.\\

\State {\it{Step 3.}} Accept $\bm{\theta}_{n+1}=\bm{\theta}_{\ast}$ with probability

$$\alpha = \min\left\lbrace \frac{p(\bm{\theta}_{\ast})\exp(\bm{\theta_{\ast}}^\intercal S_\mathbf{x})\exp(\bm{\theta}_{n}^\intercal S_\mathbf{y})q(\bm{\theta}_{n}|\bm{\theta}_{\ast})}{p(\bm{\theta}_{n})\exp(\bm{\theta}_{n}^\intercal S_\mathbf{x})\exp(\bm{\theta}_{\ast}^\intercal S_\mathbf{y})q(\bm{\theta}_{\ast}|\bm{\theta}_{n})}, 1 \right\rbrace$$ else reject (set $\bm{\theta}_{n+1}=\bm{\theta}_n$).

\end{algorithmic}
\end{algorithm}

In Step 3 in the Algorithm~\ref{DMHalg} $Z(\bm{\theta})$ get canceled, and we only need to evaluate unnormalized likelihood functions for the acceptance probability ($\alpha$) calculation. Although DMH can provide accurate inference within a reasonable computing time, DMH is still computationally expensive for large data sets \citep{park2020function}. The main computational challenge comes from the high-dimensional auxiliary variable simulations in Step 2. In what follows, we develop an indirect auxiliary variable MCMC (IAVM) that is computationally efficient and also provide accurate estimates.

\section{Bayesian Indirect Inference}

~~~~Our approach is motivated from indirect inference method \citep[cf.][]{eroux1993c,drovandi2011approximate,drovandi2015bayesian}, which uses a surrogate model for approximating the intractable true model. For this approach, we can define a binding function, which maps the parameters of the true model into the parameters of the surrogate model. 
Our approach has similarities to \cite{wood2010statistical,moores2015pre} in that we use a normal distribution as a surrogate, but our method can be more broadly applicable to multi-dimensional doubly-intractable distributions. 

\subsection{Outline}

\begin{figure*}[tt]
\begin{center}
\includegraphics[ scale = 1.4]{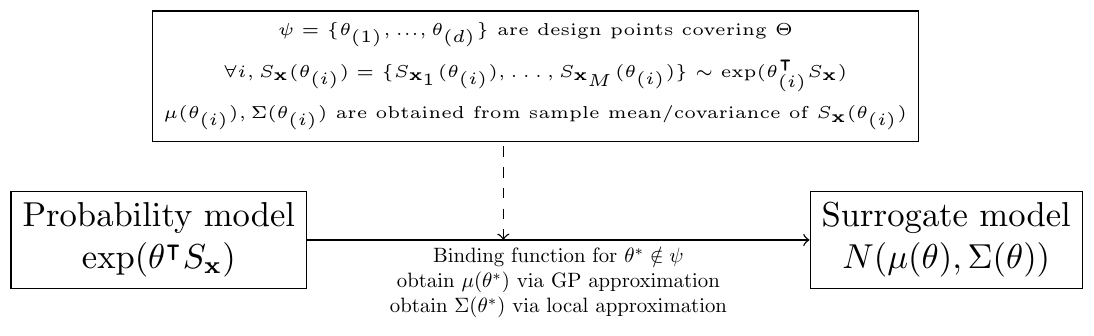}
\end{center}
\caption[]{Illustration for the precomputation step of the indirect auxiliary variable MCMC (IAVM).}
\label{IAVMfigure}
\end{figure*}

~~~~The main idea of our approach is to generate the summary statistics from the computationally cheap surrogate model, rather than to simulate this from the expensive true model. We use Gaussian processes \citep[cf.][]{rasmussen2004gaussian} as a binding function to connect the probability model and a surrogate model (see Figure~\ref{IAVMfigure}).
We begin with an outline of the indirect auxiliary variable MCMC (IAVM) as follows. \\

\noindent {\it{Step 1.}} $M$-number of summary statistics are generated from the true probability model at a set of $\bm{\theta}$ values. \\
\noindent {\it{Step 2.}} For each $\bm{\theta}$ value, the distribution of the summary statistics is approximated via a normal distribution with mean $\bm{\mu}(\bm{\theta})$ and covariance $\bm{\Sigma}(\bm{\theta})$ which are obtained from sampled summary statistics in Step 1.\\
{\it{Step 3.}} A Gaussian process model (binding function) is fit to the above sample mean, which allows for approximation of $\bm{\mu}(\bm{\theta}^\ast)$ at any $\bm{\theta}^\ast$ value.\\
{\it{Step 4.}} The Auxiliary variable MCMC algorithm is constructed for sampling from the posterior distribution of $\bm{\theta}$. For any $\bm{\theta}^\ast$, the summary statistics are simulated from a normal distribution with the mean approximated from Step 3, and the covariance chosen from $\bm{\Sigma}(\bm{\theta})$, where $\bm{\theta}$ is the closest point from $\bm{\theta}^\ast$. \\

\noindent Step 1 - 3 is the precomputation step, which can be embarrassingly parallel. We summarize the connection between the probability model and the surrogate model in Figure~\ref{IAVMfigure}. We provide details in the following section. 

\subsection{Indirect Auxiliary Variable Markov Chain Monte Carlo}

~~~~Let $\bm{\theta} \in \bm{\Theta}$ be a $p$-dimensional parameter vector. Consider $d$-number of design points, $\bm{\psi} =( \bm{\theta}_{(1)},\dots,\bm{\theta}_{(d)} )^\intercal$ which cover the important region of the parameter space. We use pseudolikelihood approximation \citep{besag1974spatial} to choose design points. We generate design points from a heavy-tailed multivariate $t$ distribution with mean at the maximum pseudolikelihood estimate (MPLE) and covariance obtained from the corresponding Hessian matrix. There can be other options for choosing design points. For example, we can use a short run of DMH or ABC as in \cite{park2020function}. For each $\bm{\theta}_{(i)}$, $M$-number of independent summary statistics $\lbrace S_{\mathbf{x}_{1}}(\bm{\theta}_{(i)}),\dots,S_{\mathbf{x}_{M}}(\bm{\theta}_{(i)})\rbrace$ are generated via MCMC algorithm  whose stationary distribution is $\exp(\bm{\theta}_{(i)}^\intercal S_\mathbf{x})$. Then, the distribution of the summary statistics is approximated via a normal distribution with mean $\bm{\mu}(\bm{\theta}_{(i)})$ and covariance $\bm{\Sigma}(\bm{\theta}_{(i)})$ which are obtained from the sampled $M$-number of summary statistics. 

Then, we construct binding functions to obtain an approximation of $\bm{\mu}(\bm{\theta}_\ast)$ for an arbitrary $\bm{\theta}_\ast$ value. We fit Gaussian process models relating the mean of the summary statistics $\bm{\mu} = (\bm{\mu}(\bm{\theta}_{(1)}),\dots,\bm{\mu}(\bm{\theta}_{(d)}) )^\intercal$ to the design points $\bm{\psi}$. Then the Gaussian process models can be written as follows. 

\begin{equation}
\bm{\mu}= \bm{\psi}\bm{\beta} + \mathbf{u}, ~~\mathbf{u} \sim N(0, \mathbf{K}),
\label{GP}
\end{equation}

\noindent where $\bm{\beta}$ is the regression parameter and $\mathbf{u}$ is a second order stationary Gaussian process. The covariance of $\mathbf{u}$ allows for a "nonparametric" non-linear trend, which is the basis for kriging and computer model emulation. For $i,j =1,\dots,d$, a symmetric and positive definite covariance function $\mathbf{K}(\bm{\theta}^{(i)},\bm{\theta}^{(j)};\sigma^2,\phi,\tau^2)$ can be defined as

\begin{equation}
\sigma^2\Big(  1 + \frac{\sqrt{3}\|\bm{\theta}^{(i)}-\bm{\theta}^{(j)}\|}{\phi} \Big) \exp{\Big( \frac{-\sqrt{3}\|\bm{\theta}^{(i)}-\bm{\theta}^{(j)}\|}{\phi}   \Big)} + \tau^2 1_{\lbrace i = j \rbrace},
\label{matern}
\end{equation}	

\noindent with partial sill $\sigma^2$, range $\phi$, and nugget $\tau^2$. We use a  Mat\'{e}rn class covariance function where the smoothness parameter is set to $3/2$, because we assume that the $\bm{\mu}(\bm{\theta})$ surface is smooth. To obtain $\bm{\mu}(\bm{\theta}_\ast)$ at some new $\bm{\theta}_{\ast}\in \Theta$, we use definitions of the conditional distributions for multivariate normal distributions to have

\begin{equation}
\begin{bmatrix}
\bm{\mu} \\
\bm{\mu}(\bm{\theta}_\ast)
\end{bmatrix}
=MVN\left( \begin{bmatrix}
     \bm{\psi}\bm{\beta} \\
     \bm{\theta}_{\ast}\bm{\beta} 
     \end{bmatrix}
   ,
  \begin{bmatrix}
    \mathbf{C} & \mathbf{c} \\
    \mathbf{c}' & \sigma^2+\tau^2
  \end{bmatrix}
  \right),
\label{GPmat}
\end{equation}

\noindent where $\mathbf{C}=\mathbf{K}(\bm{\psi},\bm{\psi};\sigma^2,\phi) \in R^{d \times d}$ and $\mathbf{c}=\mathbf{K}(\bm{\psi},\bm{\theta}^{\ast};\sigma^2,\phi) \in R^{d \times 1}$. Then, the conditional distribution of $\bm{\mu}(\bm{\theta}_\ast)$ given  observed $\bm{\mu}$ can be written as 

\begin{equation}
\bm{\mu}(\bm{\theta}_\ast)|\bm{\mu} \sim N( \bm{\theta}^{\ast}\bm{\beta}  + \mathbf{c}'\mathbf{C}^{-1}(\bm{\mu}-\bm{\psi}\bm{\beta} ), \sigma^{2}+\tau^2-\mathbf{c}'\mathbf{C}\mathbf{c}).
\label{condexp}
\end{equation}

\noindent A generalized least squares (GLS) estimator of regression parameter is $\widehat{\bm{\beta}}=(\bm{\psi}'\mathbf{C}^{-1}\bm{\psi})^{-1}\bm{\psi}'\mathbf{C}^{-1}\bm{\mu}$ for given true covariance parameters $( \sigma^{2},\phi, \tau^2 )$. Then, the best linear unbiased predictor (BLUP) for $\bm{\mu}(\bm{\theta}_\ast)$ can be obtained as 

\begin{equation}
\widehat{\bm{\mu}}(\bm{\theta}_\ast) =\bm{\theta}_{\ast}\widehat{\bm{\beta}} + \mathbf{c}'\mathbf{C}^{-1}(\bm{\mu} -\bm{\psi}\widehat{\bm{\beta}}).
\label{BLUP1}
\end{equation}

\noindent In practice, by plugging estimates of covariance parameters $( \sigma^{2},\phi,\tau^2 )$ (e.g., maximum likelihood or ordinary least squares) into the covariance $\mathbf{c},\mathbf{C}$, we can construct a GLS estimate $\widehat{\bm{\beta}}$. Using these plug-in estimates, \eqref{BLUP1} is called the empirical BLUP (EBLUP).

To obtain the covariance $\Sigma(\bm{\theta}_\ast)$ for an arbitrary $\bm{\theta}_\ast$, consider the closest design point $\bm{\theta}_{(l)}$ from a  $\bm{\theta}_\ast$ value in terms of the Euclidean distance. Then we set

\begin{equation}
\widehat{\bm{\Sigma}}(\bm{\theta}_{\ast}) = \bm{\Sigma}(\bm{\theta}_{(l)}),~~~~l = \min_{i}\|\bm{\theta}_{(i)}-\bm{\theta}_{\ast}\|.
\label{Covbinding}
\end{equation} 

\noindent With each iteration of the MCMC algorithm, we generate an auxiliary variable from a normal distribution with approximated mean $\widehat{\bm{\mu}}(\bm{\theta}_\ast)$ and covariance $\widehat{\bm{\Sigma}}(\bm{\theta}_{\ast})$. The indirect auxiliary variable MCMC (IAVM) algorithm is described in Algorithm~\ref{IAVMalg}.

\begin{algorithm}[hh]
\caption{Indirect Auxiliary Variable MCMC (IAVM) }\label{IAVMalg}
\begin{algorithmic}
\normalsize

\State \textbf{Part 1: Approximate the distribution of the summary statistics via a normal distribution.}\\

\State {\it{Step 1.}} For each $\bm{\theta}_{(i)}$, a set of $M$ independent summary statistics $\lbrace S_{\mathbf{x}_{1}}(\bm{\theta}_{(i)}),\dots,S_{\mathbf{x}_{M}}(\bm{\theta}_{(i)}) \rbrace$ are generated via MCMC algorithm  each with stationary distribution $\exp(\bm{\theta}_{(i)}^\intercal S_{\mathbf{x}})$.\\

\State {\it{Step 2.}} For each $\bm{\theta}_{(i)}$ value, the distribution of the summary statistics is approximated via a normal distribution with sample mean $\bm{\mu}(\bm{\theta}_{(i)})$ and sample covariance $\bm{\Sigma}(\bm{\theta}_{(i)})$.\\

\State {\it{Step 3.}}  Obtain hyper-parameters  $( \sigma^2, \phi, \tau^2, \bm{\beta} )$ by fitting a Gaussian process via MLE to $\lbrace (\bm{\theta}_{(1)}, \bm{\mu}(\bm{\theta}_{(1)})),\dots,(\bm{\theta}_{(d)}, \bm{\mu}(\bm{\theta}_{(d)})) \rbrace$.\\

\State \textbf{Part 2: MCMC with an indirect auxiliary variable simulation.}\\

\State Given $\bm{\theta}_{n} \in \bm{\Theta}$ at $n$th iteration, construct the next step of the algorithm as follows. \\

\State {\it{Step 4.}} Propose $\bm{\theta}_{\ast} \sim q(\cdot|\bm{\theta}_{n})$.\\

\State {\it{Step 5.}} Generate the sufficient statistics of an auxiliary variable from normal distribution: $S_\mathbf{y} \sim N( \widehat{\bm{\mu}}(\bm{\theta}_{\ast}), \widehat{\bm{\Sigma}}(\bm{\theta}_{\ast}) )$, where $\widehat{\bm{\mu}}(\bm{\theta}_{\ast}),\widehat{\bm{\Sigma}}(\bm{\theta}_{\ast})$ are obtained as in \eqref{BLUP1} and \eqref{Covbinding}.\\

\State {\it{Step 6.}} Accept $\bm{\theta}_{n+1}=\bm{\theta}_{\ast}$ with probability

$$\alpha = \min\left\lbrace \frac{p(\bm{\theta}_{\ast})\exp(\bm{\theta_{\ast}}^\intercal S_\mathbf{x})\exp(\bm{\theta}_{n}^\intercal S_\mathbf{y})q(\bm{\theta}_{n}|\bm{\theta}_{\ast})}{p(\bm{\theta}_{n})\exp(\bm{\theta}_{n}^\intercal S_\mathbf{x})\exp(\bm{\theta}_{\ast}^\intercal S_\mathbf{y})q(\bm{\theta}_{\ast}|\bm{\theta}_{n})}, 1 \right\rbrace$$ else reject (set $\bm{\theta}_{n+1}=\bm{\theta}_n$).

\end{algorithmic}
\end{algorithm}

The indirect auxiliary variable MCMC (IAVM) can dramatically reduce the computational cost by simulating the summary statistics from a normal distribution (Step 5 in Algorithm~\ref{IAVMalg}). This is much cheaper than simulating the summary statistics from a true probability model as in DMH (Step 2 in Algorithm~\ref{DMHalg}). Furthermore, in the precomputation step, we can use parallel computation because generating a set of auxiliary variables for each design point is independent (Step 1 in Algorithm~\ref{IAVMalg}). In Section 4, we implement this through {\tt OpenMP } across independent processors. 

In the supplementary material, we provide theoretical justification for IAVM. When the true distribution of the summary statistics $\exp(\bm{\theta}_{(i)}^\intercal S_\mathbf{x})$ is normal, the Markov chain samples from IAVM will be close to the target distribution $\pi(\bm{\theta}|\mathbf{x})$ with increasing $M$ and $d$. In practice, with finite $M$ and $d$, IAVM is asymptotically inexact. We note that asymptotically exact approaches for doubly-intractable distributions are possible only for a few special cases, for instance, in small Ising models. In the case of large social network examples with multiple model parameters, no existing work provides exact estimates. Therefore, in Section 4, we compared IAVM with the existing inexact approach and found that IAVM can provide reasonable approximation at a far lower computational cost.

\section{Simulated and Real Data Examples}

~~~~We study our approach to the Ising model and two large social network examples. To show the computational efficiency of our approach, we compare indirect auxiliary variable MCMC (IAVM) with double Metropolis-Hastings (DMH). DMH was found to be the only applicable option to computationally expensive problems when exact approaches are not possible \citep{park2018bayesian}. For social network examples, IAVM shows a dramatic computational gain over DMH.

The code is implemented in {\tt R} and {\tt C++}. We estimate the hyper-parameters $( \sigma^2, \phi, \tau^2, \bm{\beta} )$ of Gaussian process models using the \texttt{DiceKriging} package. We obtain point estimates through simple means of the entire posterior sample without thining or burn-in. The calculation of Effective Sample Size (ESS) follows \cite{robert2013monte}. We calculate the highest posterior density (HPD) by using \texttt{coda} package in {\tt R}. All the code was implemented on dual 10 core Xeon E5-2680 processors on the Penn State high performance computing cluster.

\subsection{The Ising Model} 

~~~~The Ising model is a non-Gaussian Markov random field, and is used to describe spatial association on an $m$ by $n$ lattice. Consider the observed lattice  $\mathbf{x} \in R^{m \times n}$, which has binary values $x_{i,j}=\lbrace -1,1 \rbrace$. The model is

 \begin{equation}
L(\theta|\mathbf{x})=\frac{1}{Z(\theta)}\exp\left\lbrace \theta S_\mathbf{x} \right\rbrace, 
\label{Isingeq}
\end{equation}
\begin{gather*}
S_\mathbf{x}=\sum_{i=1}^{m}\sum_{j=1}^{n-1}x_{i,j}x_{i,j+1} + \sum_{i=1}^{m-1}\sum_{j=1}^{n}x_{i,j}x_{i+1,j}.
\end{gather*}

\noindent The sufficient statistic $S_\mathbf{x}$ imposes spatial dependence; the larger value indicates the stronger interactions. Here, summation over all $2^{mn}$ possible lattice configurations is required for $Z(\bm{\theta})$ calculation, which is intractable. We simulate 100$\times$100 lattice data using perfect sampling \citep{propp1996exact} with $\theta=0.3$.

\begin{figure*}[tt]
\begin{center}
\includegraphics[ scale = 0.6]{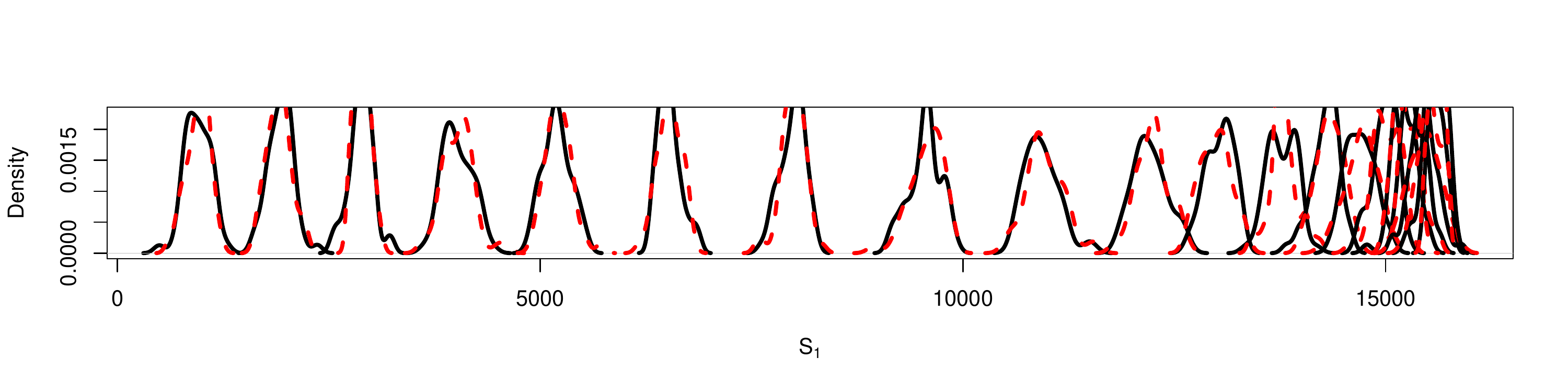}
\end{center}
\caption[]{Comparison of the distribution of the summary statistics generated from a standard Gibbs sampler (solid lines) with a normal approximation (dashed lines) in the Ising example. Both are simulated 50 times at each design point.}
\label{Isingfigure}
\end{figure*}

In IAVM, we generated $d=20$ number of design points uniformly over the prior region [0,1]. We used parallel computation to generate $M=50$ number of auxiliary variables from the probability model at each design point. The parallel computing was implemented through {\tt OpenMP } across 20 independent processors. We can choose $M$ by comparing the true distribution of the summary statistics with a normal approximation at each design point. For each design point, we generate $M=50$ number of summary statistics from the true model. Then we compare the distribution of the summary statistics with a normal surrogate. Figure~\ref{Isingfigure} indicates agreement between these distributions. We used a normal proposal with a standard deviation of 0.01 for all algorithms. Compared to other approaches, the exchange algorithm \citep{murray2006} is asymptotically exact because the auxiliary variable is exactly generated from the true probability model. All the algorithms were run until the Monte Carlo standard errors are at or below 0.001.

\begin{table}[tt]
\centering
\begin{tabular}{cc}
  \hline
  & $\theta =0.3$ \\
  \hline
$\mathbf{DMH}$ \\
Mean & 0.30    \\
95\%HPD & (0.29, 0.31)  \\
ESS &  1049.38  \\
Time(second) & 113.25 \\
ESS/Time & 9.27 \\
  \hline
$\mathbf{IAVM}$   \\
Mean & 0.30 \\
95\%HPD & (0.29, 0.31) \\
ESS & 1206.67 \\
Time(second) & 47.89\\
ESS/Time & 25.20  \\
  \hline  
$\mathbf{Exchange}$ & \\
Mean & 0.30 \\
95\%HPD & (0.29, 0.31)  \\
ESS & 1103.37 \\
Time(second) & 1483.39 \\
ESS/Time & 0.74\\
\hline
\end{tabular}
\caption{Inference results for the Ising model simulation study. 10,000 MCMC samples are generated from each algorithm.}
\label{Isingtable} 
\end{table}

For this small Ising model example, IAVM is about two times faster than DMH. Including precomputing time, IAVM takes about 1 minute, while DMH takes about 2 minutes. We observe that the exchange algorithm takes about 30 minutes because perfect sampling takes longer to achieve coalescence, even for this small lattice problem. Table~\ref{Isingtable} shows that the estimates from all algorithms are comparable. We also calculate effective sample size (ESS) to approximates the number of independent samples that correspond to the number of correlated samples from the Markov chain; ESS for algorithms are similar. In summary,  we observed that IAVM provides an accurate estimate, and is faster than DMH, but does not show significant differences. This is because auxiliary variable simulations are not that expensive in this example. In the following section, we study large network models and show how IAVM has the potential for greater gains for more challenging problems.

\subsection{The International E-road Network}

~~~~Exponential random graph models (ERGM) \citep{robins2007introduction} are widely used to describe relationships among nodes in networks. In this manuscript, we consider the undirected ERGM with $n$ nodes. Consider a network matrix $\mathbf{x} \in R^{n \times n}$. For all $i \neq j$, $x_{i,j}=1$ if the $i$th node and $j$th node are connected, otherwise $x_{i,j}=0$ and $x_{i,i}$ is defined as 0. Calculation of the $Z(\bm{\theta})$ requires summation over all $2^{n(n-1)/2}$ network configurations, which is intractable. Here, we study the international E-road network \citep{vsubelj2011robust,nr}, which describes connections among 1177 European cities. The international E-road network data set may
be downloaded from their network repository (http://networkrepository.com/road.php). Consider the ERGM, where the probability model is

\begin{equation}
L(\bm{\theta}|\mathbf{x})=\frac{1}{Z(\bm{\theta})}\exp\left\lbrace \theta_{1}S^1_{\mathbf{x}} + \theta_{2}S^2_{\mathbf{x}} \right\rbrace,
\label{eroad}
\end{equation}
\begin{gather*}
S^1_{\mathbf{x}}=\sum_{i=1}^{n}{x_{i+} \choose 1} ~S^2_{\mathbf{x}}=e^{\tau_s}\sum_{k=1}^{n-2}\left\lbrace 1-(1-e^{-\tau_s})^{k} \right\rbrace ESP_{k}(\mathbf{x}).
\end{gather*}

\noindent Here $S^1_{\mathbf{x}}$ is the number of edges and $S^2_{\mathbf{x}}$ denotes the geometrically weighted edge-wise shared partnership statistic (GWESP) \citep{hunter2006inference}. $ESP_{k}(\mathbf{x})$ counts the number of connected pairs $(i,j)$, which have $k$ common neighbors. By placing geometric weights $\tau_s=0.25$ on the edges with higher transitivities, GWESP describes edge-wise shared partnership. We used uniform priors on $\bm{\theta}$. The uniform prior is centered around the MPLE with a width of 10 standard deviations. For this ERGM, auxiliary variables can be generated from the probability model through 1 cycle of Gibbs sampling. This means that the number of Metropolis-Hastings updates in Step 2 of Algorithm~\ref{DMHalg} is equal to $n(n-1)/2$, where $n=1177$. With each iteration, the $ij$th element of a data matrix is randomly chosen, and $x_{i,j}$ is set to 0 or 1 following the full conditional probabilities.

\begin{figure*}[tt]
\begin{center}
\includegraphics[ scale = 0.5]{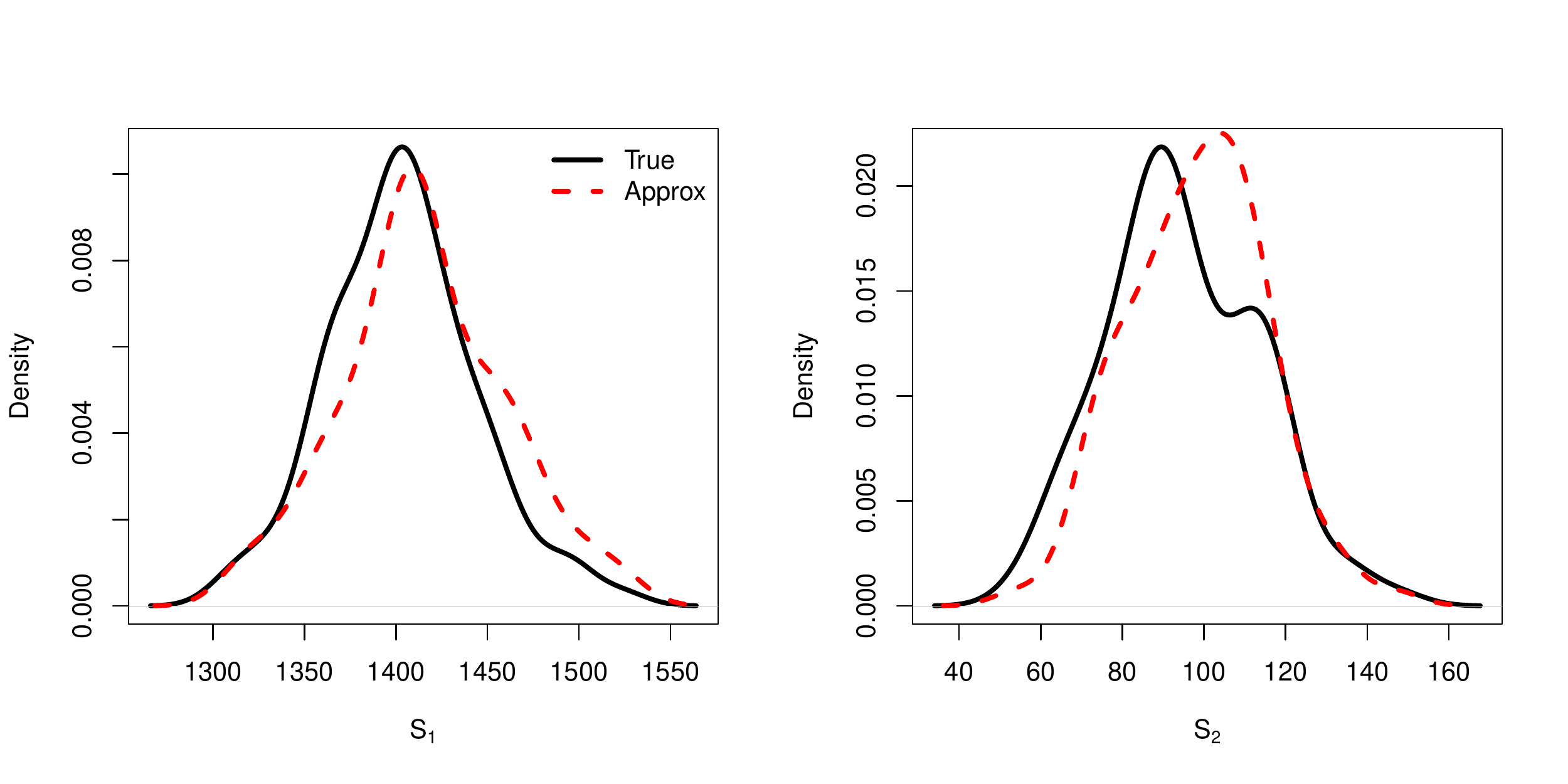}
\end{center}
\caption[]{Comparison of the distribution of the summary statistics generated from a standard Gibbs sampler (solid lines) with a normal approximation (dashed lines) in the E-road network example. Both are simulated 100 times at MPLE.}
\label{eroadfigure}
\end{figure*}

In IAVM, we generated $d=400$ number of design points via a pseudolikelihood approximation, as before. We used parallel computation to generate $M=50$ number of auxiliary variables from the probability model at each design point. The performance of a normal sampler can be checked by comparing it with the true distribution of the summary statistics. For a given MPLE, we simulate a set of summary statistics from the true probability model. Then, we can compare a normal approximation with the true distribution of the summary statistics. Figure~\ref{eroadfigure} indicates that a normal approximation to the true distribution of the summary statistics works well with $d=400$ and $M=50$. We used a multivariate normal proposal for both DMH and IAVM. The covariance matrix for the proposal is obtained from the inverse of the negative Hessian matrix at the MPLE. Unlike the Ising model example, exact approaches are infeasible for this large social network example. We run both DMH and IAVM were run until the Monte Carlo standard errors are at or below 0.001.

\begin{table}[tt]
\centering
\begin{tabular}{ccc}
  \hline
  & $\theta_{1}$ & $\theta_{2}$ \\
  \hline
$\mathbf{DMH}$ \\
Mean & -6.23  & 0.89  \\
95\%HPD & (-6.29, -6.18) & (0.77, 1.02) \\
ESS &  1717.63 & 868.52  \\
Time(hour) & 28.67 \\
minESS/Time & 30.29 \\
  \hline
$\mathbf{IAVM}$  & \\
Mean & -6.24 & 0.89\\
95\%HPD & (-6.29, -6.18) & (0.78, 1.00) \\
ESS & 1803.72 & 983.37\\
Time(hour) & 1.93\\
minESS/Time & 509.52  \\
\hline
\end{tabular}
\caption{Inference results for the E-road network example. 25,000 MCMC samples are generated from each algorithm.}
\label{eroadtable} 
\end{table}

For this large network example ($n=1177$), IAVM outperforms DMH. DMH takes about 28 hours, while IAVM takes only about 2 hours, including precomputing time. This is because auxiliary variable simulations are expensive in this example. Table~\ref{eroadtable} indicates that the estimates from both algorithms are similar. 
We observe that IAVM shows larger ESS than DMH, but does not show big differences. 

\begin{table*}[tt]
\caption{Results for $\theta_{2}$ for the E-road network example across different choices of $M$ and $d$.}
\centering
\begin{tabular}{ccccccccccc}
  \hline
$\theta_{2}$ & $M$ & $d$ & Mean & 95\%HPD  & ESS & Time(hour) & ESS/Time\\
  \hline
$\mathbf{DMH}$ & NA& NA & 0.89 & (0.77, 1.02)  & 868.52 & 28.67 & 30.29  \\
$\mathbf{IAVM}$ & 100 & 800 & 0.88 & (0.77, 1.00)  &  926.68 & 7.50 & 123.56\\
& 100 & 400 & 0.88 & (0.76, 1.01)  & 789.60 & 3.88 & 203.51\\
& 100 & 200 & 0.89 & (0.77, 1.02)  & 764.99 & 1.95 & 392.30\\
& 50 & 800 & 0.88 & (0.76, 1.00)   & 852.17 & 3.89 & 219.07 \\
& 50 & 400 & 0.89 & (0.78, 1.00)  & 983.37  & 1.93 & 509.52\\
& 50 & 200 & 0.89 & (0.77, 1.00)  & 901.18 & 1.01 & 892.26 \\
   \hline
\end{tabular}
\label{eroadMd}
\end{table*}

To validate our method, we investigate the performance of IAVM for different combinations of $M$ and $d$. The remaining settings for both algorithms are the same as before. Here, we only provide the results for $\theta_{2}$ because similar results are observed for the other parameter. We observe that IAVM can provide accurate inference results across the different choices of $M$ and $d$ (Table~\ref{eroadMd}). Naturally, computing time increases as we use larger $M$ and $d$. In summary, IAVM is much faster and provides reasonable inference results.

\subsection{A Faux Magnolia High School Network}

~~~~In this section, we study the Faux Magnolia high school network \citep{resnick1997protecting}, which describes an in-school friendship network among 1461 students. There are two node attributes in this model: (1) grade (7-12), and (2) sex (male, female). Consider the undirected ERGM, where the likelihood function is

\begin{equation}
L(\theta|\mathbf{x})=\frac{1}{Z(\theta)}\exp\left\lbrace \sum_{l=1}^{10} \theta_{l}S^l_{\mathbf{x}}\right\rbrace,
\label{emonmodel}
\end{equation}
\begin{gather*}
S^1_{\mathbf{x}}=\sum_{i=1}^{n}{x_{i+} \choose 1},~~~S^2_{\mathbf{x}}=\sum_{i<j}x_{i,j}(1\lbrace  \mbox{grade}_i=7 \rbrace+1\lbrace \mbox{grade}_j=7 \rbrace)\\
S^3_{\mathbf{x}}=\sum_{i<j}x_{i,j}(1\lbrace  \mbox{grade}_i=8 \rbrace+1\lbrace \mbox{grade}_j=8 \rbrace),~~~S^4_{\mathbf{x}}=\sum_{i<j}x_{i,j}(1\lbrace  \mbox{grade}_i=9 \rbrace+1\lbrace \mbox{grade}_j=9 \rbrace)\\
S^5_{\mathbf{x}}=\sum_{i<j}x_{i,j}(1\lbrace  \mbox{grade}_i=10 \rbrace+1\lbrace \mbox{grade}_j=10 \rbrace),~~~S^6_{\mathbf{x}}=\sum_{i<j}x_{i,j}(1\lbrace  \mbox{grade}_i=11 \rbrace+1\lbrace \mbox{grade}_j=11 \rbrace)\\
S^7_{\mathbf{x}}=\sum_{i<j}x_{i,j}(1\lbrace  \mbox{grade}_i=12 \rbrace+1\lbrace \mbox{grade}_j=12 \rbrace),~~~S^8_{\mathbf{x}}=\sum_{i<j}x_{i,j}(1\lbrace  \mbox{sex}_i=\mbox{male} \rbrace+1\lbrace \mbox{sex}_j=\mbox{male} \rbrace)\\
S^9_{\mathbf{x}}=e^{\tau_d}\sum_{k=1}^{n-1}\left\lbrace 1-(1-e^{-\tau_d})^{k} \right\rbrace D_{k}(\mathbf{x}),~~~S^{10}_{\mathbf{x}}=e^{\tau_s}\sum_{k=1}^{n-2}\left\lbrace 1-(1-e^{-\tau_s})^{k} \right\rbrace ESP_{k}(\mathbf{x})
\end{gather*}

\noindent The sufficient statistics are $S^1_{\mathbf{x}}$ (the number of edges), $S^2_{\mathbf{x}}-S^7_{\mathbf{x}}$ (node factor for grade), $S^8_{\mathbf{x}}$ (node factor for sex), $S^9_{\mathbf{x}}$ (geometrically weighted degree (GWD) statistics), and $S^{10}_{\mathbf{x}}$ (GWESP). For the GWD statistics, $D_{k}(\mathbf{x})$ counts the number of nodes which have $k$ relationships. For GWD and GWESP statistics, we use $\tau_d=\tau_s=0.25$ to place geometric weights. As in the previous example, we used uniform priors on $\bm{\theta}$; centered around the MPLE with a width of 10 standard deviations. For this model, auxiliary variables are generated from the probability model via 1 cycle of Gibbs sampler.

\begin{figure*}[tt]
\begin{center}
\includegraphics[ scale = 0.5]{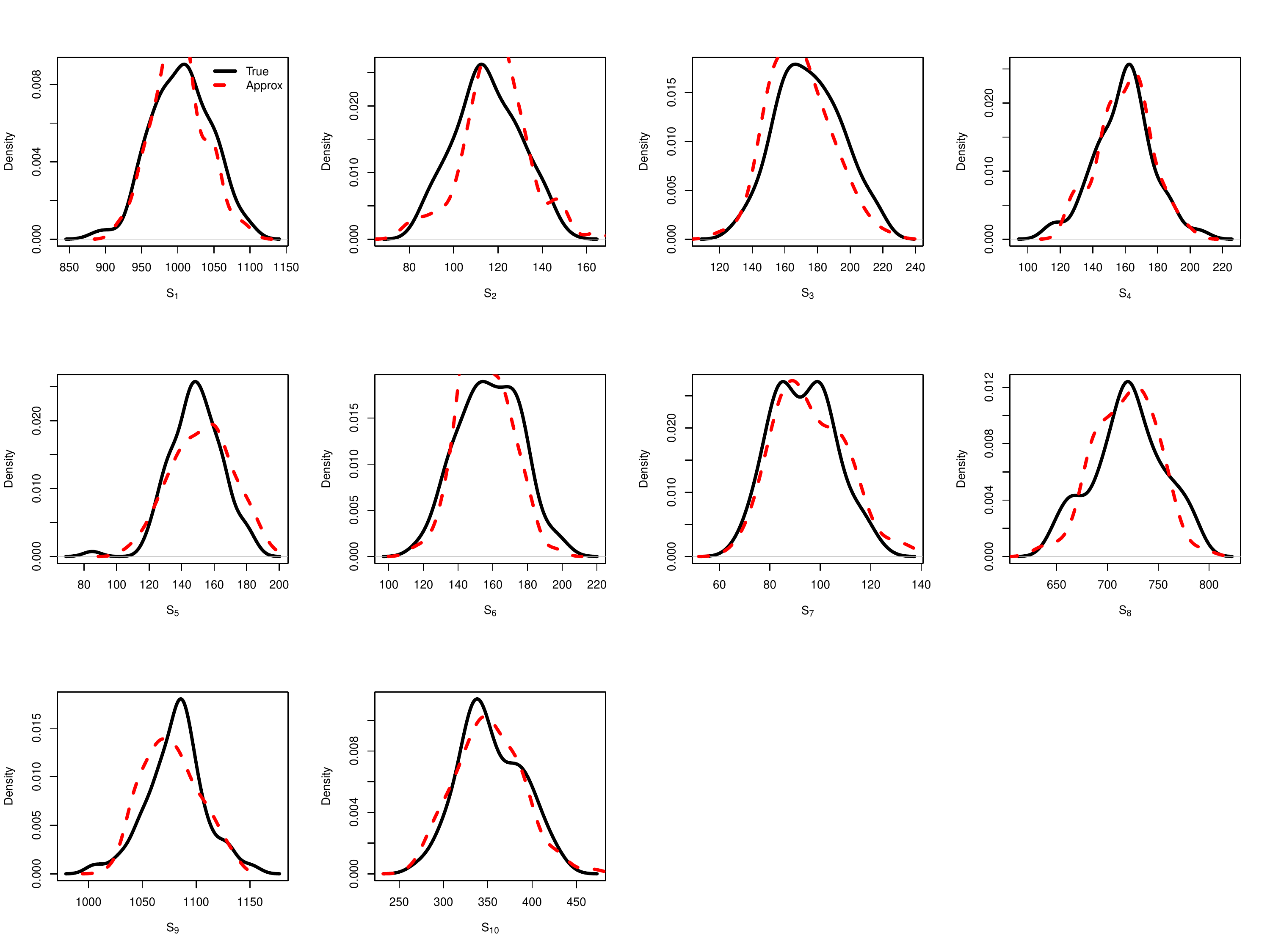}
\end{center}
\caption[]{Comparison of the distribution of the summary statistics generated from a standard Gibbs sampler (solid lines) with a normal approximation (dashed lines) in the Faux Magnolia high school network example. Both are simulated 100 times at MPLE.}
\label{fauxmagnoliafigure}
\end{figure*}

For IAVM, we generated $M=50$ number of auxiliary variables from the probability model at $d=400$ number of design points by using parallel computation. As described in Section 3.3, design points are generated through a pseudolikelihood approximation. Figure~\ref{fauxmagnoliafigure} shows that a normal approximation is well matched to the true distribution of the summary statistics for $d=400$ and $M=50$. For both DMH and IAVM, we used a multivariate normal proposal, with the covariance obtained as in the previous example. Both algorithms were run until the Monte Carlo standard errors are at or below 0.005.

\begin{table*}[tt]
\caption{Inference results for a Faux Magnolia high school network example. 80,000 MCMC samples are generated from each algorithm.}
\centering
\begin{tabular}{cccccc}
  \hline
  & $\theta_{1}$ & $\theta_{2}$ & $\theta_{3}$ & $\theta_{4}$ & $\theta_{5}$ \\
  \hline
$\mathbf{DMH}$ & \\
Mean & -9.06 & 3.03 & 3.03 & 2.52 & 2.62\\
95\%HPD & (-9.30, -8.80) & (2.76, 3.30) & (2.77, 3.27) & (2.28, 2.75) & (2.38, 2.86)\\
ESS & 1142.73  & 1208.12 & 1213.49 & 1220.50 & 1227.32\\
  \hline
$\mathbf{IAVM}$ & \\
Mean & -9.07 & 3.03 & 3.05 & 2.52 & 2.62\\
95\%HPD & (-9.31, -8.82) & (2.76, 3.30) & (2.81, 3.30) & (2.26, 2.75) & (2.37, 2.88) \\
ESS & 1142.19 &  1201.98 &  1238.93 & 1237.14 & 1180.59\\
\hline
  & $\theta_{6}$ & $\theta_{7}$ & $\theta_{8}$ & $\theta_{9}$ & $\theta_{10}$ \\
  \hline
$\mathbf{DMH}$ & \\
Mean & 2.79 & 2.90 & 0.78 & -0.15 & 1.82 \\
95\%HPD &  (2.54, 3.02) & (2.63, 3.20) & (0.64, 0.94) & (-0.32, 0.04) & (1.69, 1.95) \\
ESS & 1217.96  & 1210.91 & 1236.02 & 1055.75 & 979.54\\
Time(hour) & 338.42\\
minESS/Time & 2.89 \\
  \hline
$\mathbf{IAVM}$ & \\
Mean & 2.80 & 2.92 & 0.78 & -0.14 & 1.81\\
95\%HPD & (2.55, 3.05) & (2.64, 3.19) & (0.62, 0.94) & (-0.33, 0.06) & (1.68, 1.92) \\
ESS & 1194.93 & 1234.12 & 1203.94 & 1041.45 &  970.27 \\
Time(hour) & 5.25 \\
minESS/Time & 184.81 \\
\hline
\end{tabular}
\label{fauxmagnolia} 
\end{table*}

The IAVM shows a dramatic speed-up over DMH methods. While DMH takes about 2 weeks for model fitting, IAVM only takes about 5 hours, including precomputing time. This is because high-dimensional auxiliary variable simulations are extremely expensive for this large network model. Table~\ref{fauxmagnolia} shows that posterior estimates from DMH and IAVM are similar to each other. We observe that ESS for both algorithms are similar. In summary, IAVM provides accurate inference results within a much shorter time. IAVM has significant computational advantages over DMH for doubly-intractable distributions with large data sets and moderate dimensional parameter space. 

\begin{table*}[tt]
\caption{Results for $\theta_{10}$ for a Faux Magnolia high school network example across different choices of $M$ and $d$.}
\centering
\begin{tabular}{cccccccccc}
  \hline
$\theta_{10}$ & $M$ & $d$ & Mean & 95\%HPD & ESS & Time(hour) & ESS/Time\\
  \hline
$\mathbf{DMH}$ & NA& NA & 1.82 & (1.69, 1.95) & 975.54 & 338.42 &  2.89\\
$\mathbf{IAVM}$ & 100 & 800 & 1.82 & (1.69, 1.96) & 944.54 &  20.92 &  45.15\\
& 100 & 400 & 1.82  & (1.70, 1.93) & 968.68 & 10.51 & 92.13\\
& 100 & 200 & 1.82 & (1.70, 1.95) & 990.50 & 5.40 & 183.46\\
 & 50 & 800 & 1.82 & (1.69, 1.93) & 986.40 & 11.26 & 87.59 \\
 & 50 & 400 & 1.81 & (1.68, 1.92) & 970.27 & 5.25 & 184.81 \\
 & 50 & 200 & 1.83  & (1.70, 1.95)  & 968.67  & 2.93 & 330.41 \\
   \hline
\end{tabular}
\label{fauxmagnoliaMd}
\end{table*}

We study the performance of IAVM for different choices of $M$ and $d$. The rest of the settings for all algorithms are the same. Table~\ref{fauxmagnoliaMd} shows that IAVM can recover posterior distribution well, compared to DMH across different combinations of $M$ and $d$. This fact demonstrates that IAVM is robust for the choice of $M$ and $d$ even for models with 10-dimensional parameter space. 
This study highlights the fact that IAVM can provide accurate results for large networks up to 10-dimensional model parameters within a reasonably short time.

\section{Discussion}

~~~~We have proposed an efficient indirect auxiliary variable MCMC (IAVM) algorithm by replacing an expensive auxiliary variable simulation from a true probability model with a fast simulation from a surrogate normal distribution. For any $\bm{\theta}_{\ast}$ values, we interpolate the mean of the normal distribution by using the Gaussian process approximation. Our study to social network applications shows that IAVM provides similar results to DMH and dramatically reduces computing costs. We observe that our method can reduce computing time from about 2 weeks to only about 5 hours. Considering that no existing approaches are feasible for doubly intractable distributions with moderately large dimensional $\bm{\theta}$ for large $\mathbf{x}$, IAVM has significant gains for more challenging cases.

The main computational benefits of our method come from using parallel computation in the precomputation step (Step 1 in Algorithm~\ref{IAVMalg}) by generating a set of auxiliary variables from design points. Computational costs can be reduced by a factor corresponding to the number of available cores, which can improve the scalability of the algorithm. Our work is motivated by recently developed precomputation approaches for intractable normalizing function problems. For instance, \cite{boland2018efficient,park2020function} construct the importance sampling estimates in parallel in their precomputation step. \cite{moores2015pre} develops an efficient preprocessing approach in the approximate Bayesian computation context. Given the increasing availability of scientific computing, this will be of particular interest.

We assume that the true distribution of the summary statistics is unimodal as in \cite{moores2015pre,cabras2014quasi}. As our examples in Section 4 illustrate, such an assumption works well for many interesting applications. We note that degeneracy in ERGM can pose challenges on MCMC simulations; the simulated networks are fully connected (complete) or entirely unconnected (empty). In this case, there might be some irregularities in the distribution of the summary statistics (e.g., multimodalities). Since the observed network is less likely to be complete or empty, in the Bayesian analysis, we can specify a prior to the non-degeneracy region with some preliminary exploration of the parameter space. For example, \cite{jin2013bayesian} uses ABC to rule out the degeneracy region before implementing their Monte Carlo algorithm. 

With appropriate choice of summary statistics such as Ripley's K-function \citep{ripley1976second}, we can apply IAVM to point process models. The method we propose in this manuscript is effective for a moderate parameter dimension ($p$) with a large sample size ($n$). For these settings, IAVM would perform better than DMH. However, with increasing parameter dimensions, the number of design points for Gaussian process models should be increased, which slows computing. We note that inference for doubly intractable distributions with large $n$ and high-dimensional $p$ is challenging, and remains an open question. Our method can be applicable to discrete parameter spaces. Instead of using continuous proposals such as symmetric normal distribution, we can use an irreducible transition matrix to propose parameters from the discrete state spaces. Then for an arbitrary $\bm{\theta}^\ast$ value, we can evaluate $\widehat{\bm{\mu}}(\bm{\theta}^\ast)$ through Gaussian process approximation. Once we have a moderate number of parameters (up to 10), IAVM will be computationally efficient than the existing method. This might be useful for sampling from discrete parameter spaces, such as the interaction radius parameter at different levels in the Strauss point process \citep{strauss1975model}.

We note that our methods require low-dimensional summary statistics, which is available for many cases, for example, in social networks and image analysis. However, applying our method to doubly-intractable distributions without such summary statistics \citep[cf.][]{goldstein2014attraction} is not trivial. This is because direct approximation to the distribution of the high-dimensional auxiliary variable is challenging. Developing extensions of our method to such models will be an interesting avenue for future research.


\section*{Acknowledgement}
Jaewoo Park is partially supported by the Yonsei University Research Fund of 2019-22-0194 and the National Research Foundation of Korea (NRF-2020R1C1C1A01003868). The author is grateful to the anonymous reviewers for their careful reading and valuable comments.

\section*{Disclosure statement}
No potential conflict of interest was reported by the author.
\clearpage

\appendix
\begin{center}
\title{\LARGE\bf Supplementary Material for Bayesian Indirect Inference for Models with Intractable Normalizing Functions}\\~\\
\author{\Large{Jaewoo Park}}
\end{center}

\section{Theoretical Justifications}

~~~~For IAVM, we study the approximation error in terms of total variation distance \citep{atchade2014bayesian, alquier2014noisy}. Consider a target distribution $\pi(\bm{\theta}|\mathbf{x})$. By generating the sufficient statistics of an auxiliary variable from the probability model $\exp(\bm{\theta}^\intercal S_{\mathbf{y}})$, we can construct the Markov transition kernel $\mathbf{T}$. By replacing an auxiliary variable simulation from $\exp(\bm{\theta}^\intercal S_{\mathbf{y}})$ with a simulation from a surrogate normal distribution $g(S_\mathbf{y}|\bm{\theta})$ with mean $\mu(\bm{\theta})$ and covariance $\Sigma(\bm{\theta})$, we can construct the indirect Markov transition kernel $\mathbf{T}_{I}$, the stationary distribution of which is $\pi_{I}(\bm{\theta}|\mathbf{x})$. In practice, we approximate the surrogate model parameters via sample mean and sample covariance obtained from $M$-number of auxiliary variable samples for each $\bm{\theta}_{(i)}$ for $i=1,...,d$. Then, we can obtain the approximated indirect Markov transition kernel $\widehat{\mathbf{T}}_{I}$ whose stationary distribution is $\widehat{\pi}_{I}(\bm{\theta}|\mathbf{x})$. We make the following assumptions. 

\begin{assumption}
\label{assumption1}
$\exists$ constants $k_{h}$,$K_{h}$ s.t. $k_{h} \leq \exp(\bm{\theta}^\intercal S_\mathbf{x}) \leq K_{h}$.
\end{assumption}
\begin{assumption}
\label{assumption2}
$\exists$ constant $c_{g}>1$ s.t. $1/c_{g} \leq g(S_\mathbf{x}|\bm{\theta}) \leq c_{g}$.
\end{assumption}
\begin{assumption}
\label{assumption3}
Sample space $\mathcal{X}$ is finite
\end{assumption}
\begin{assumption}
\label{assumption4}
Parameter space $\bm{\Theta}$ is compact.
\end{assumption}

In many applications, it is reasonable to assume the parameter space is compact and the sample space is finite. In these cases, assumptions \ref{assumption1} to \ref{assumption2} are easily checked. These assumptions hold for examples in Section 4. Theorem~\ref{IAVMproof} measures the total variation distance between the target posterior distribution and the approximated indirect Markov transition kernel.

\begin{theorem}
\label{IAVMproof}
Consider Markov transition kernel $\widehat{\mathbf{T}}_{I}$ constructed by generating the summary statistics of an auxiliary variable from the normal distribution with mean $\widehat{\mu}(\bm{\theta})$ and covariance $\widehat{\Sigma}(\bm{\theta})$. Suppose Assumptions \ref{assumption1} to \ref{assumption4} hold. Then, $\| \pi(\cdot) - \widehat{\mathbf{T}}^{n}_{I}(\bm{\theta}_{0},\cdot)  \|_{TV}  \leq \rho^{n} + U +  \epsilon(M) + \epsilon(d)$ for a bounded constant $U$ and $0<\rho<1$. A bounded constant $U$ becomes $0$ when the true distribution of the summary statistics is Gaussian.
\end{theorem}

According to the theorem, the Markov chain samples from IAVM will be close to the target distribution $\pi(\bm{\theta}|\mathbf{x})$ up to a constant $U$, as the sample size for auxiliary variables ($M$) and the number of design points ($d$) are increased. When the true distribution of the summary statistics is normal, with infinitely large $M$ and $d$, the stationary distribution of IAVM converges to the target distribution. In practice, with finite $M$ and $d$, IAVM is asymptotically inexact.

\section{Proof of Theorem 1}

\indent Let $\bm{\theta} \in R^{p}$, and $S_{\mathbf{x}}$ and $S_{\mathbf{y}}$ are the $p$-dimensional sufficient statistics for data $\mathbf{x}$ and auxiliary variable $\mathbf{y}$ respectively. Consider a marginal target distribution $\pi(\bm{\theta}|\mathbf{x})$, where the sufficient statistics of an auxiliary variable is generated from the probability model $h(S_{\mathbf{y}}|\bm{\theta_{\ast}})$. The acceptance probability of which is 

\begin{equation}
\alpha(\bm{\theta},\bm{\theta_{\ast}}) = \min\left\lbrace \frac{p(\bm{\theta}_{\ast})\exp(\bm{\theta_{\ast}}^\intercal S_\mathbf{x})\exp(\bm{\theta}_{n}^\intercal S_\mathbf{y})q(\bm{\theta}_{n}|\bm{\theta}_{\ast})}{p(\bm{\theta}_{n})\exp(\bm{\theta}_{n}^\intercal S_\mathbf{x})\exp(\bm{\theta}_{\ast}^\intercal S_\mathbf{y})q(\bm{\theta}_{\ast}|\bm{\theta}_{n})}, 1 \right\rbrace.
\end{equation}
\noindent For measurable subset $A$ of $\bm{\Theta}$. we can define Markov transition kernel $\mathbf{T}$ as follows
\begin{equation}
\begin{split}
\mathbf{T}(\bm{\theta},A) & = 
\int_{A\times \mathcal{X}} \alpha(\bm{\theta},\bm{\theta_{\ast}})q(\bm{\theta},d\bm{\theta_{\ast}})h(dS_\mathbf{y}|\bm{\theta_{\ast}})\\
& + 1_{A}(\bm{\theta})\int_{\bm{\Theta} \times \mathcal{X}} [1-\alpha(\bm{\theta},\bm{\theta_{\ast}})]q(\bm{\theta},d\bm{\theta_{\ast}})h(dS_\mathbf{y}|\bm{\theta_{\ast}}).
\end{split}
\label{exactKernel}
\end{equation}

\noindent By replacing an auxiliary variable simulation from the model with a simulation from a multivariate normal distribution $g(S_\mathbf{y}|\bm{\theta_{\ast}})$, we can construct an indirect Markov transition kernel $\mathbf{T}_{I}$ as
\begin{equation}
\begin{split}
\mathbf{T}_{I}(\bm{\theta},A)   & = 
\int_{A\times \mathcal{X}} \alpha(\bm{\theta},\bm{\theta_{\ast}})q(\bm{\theta},d\bm{\theta_{\ast}})g(dS_\mathbf{y}|\bm{\theta_{\ast}})\\
& + 1_{A}(\bm{\theta})\int_{\bm{\Theta} \times \mathcal{X}} [1-\alpha(\bm{\theta},\bm{\theta_{\ast}})]q(\bm{\theta},d\bm{\theta_{\ast}})g(dS_\mathbf{y}|\bm{\theta_{\ast}}).
\end{split}
\label{gaussKernel}
\end{equation}

\noindent Since mean $\bm{\mu}(\bm{\theta})$ and covariance $\bm{\Sigma}(\bm{\theta})$ for $g(S_\mathbf{y}|\bm{\theta_{\ast}})$ are unknowin in practice, we approximate these parameters via sample mean $\widehat{\bm{\mu}}(\bm{\theta})$ and sample covariance $\widehat{\bm{\Sigma}}(\bm{\theta})$ obtained from $M$-number of auxiliary variable samples for each $\bm{\theta}_{(i)}$ for $i=1,...,d$. Then $g(S_\mathbf{y}|\bm{\theta_{\ast}})$ is approximated via $\widehat{g}(S_\mathbf{y}|\bm{\theta_{\ast}})$ and the corresponding transition kernel is 
\begin{equation}
\begin{split}
\widehat{\mathbf{T}}_{I}(\bm{\theta},A) &  = 
\int_{A\times \mathcal{X}} \alpha(\bm{\theta},\bm{\theta_{\ast}})q(\bm{\theta},d\bm{\theta_{\ast}})\widehat{g}(dS_\mathbf{y}|\bm{\theta_{\ast}}) \\
& + 1_{A}(\bm{\theta})\int_{\bm{\Theta} \times \mathcal{X}} [1-\alpha(\bm{\theta},\bm{\theta_{\ast}})]q(\bm{\theta},d\bm{\theta_{\ast}})\widehat{g}(dS_\mathbf{y}|\bm{\theta_{\ast}}).
\end{split}
\label{approxgaussKernel}
\end{equation}

\noindent We first show the existence of unique invariant distributions $\pi$, $\pi_{I}$ and $\widehat{\pi}_{I}$ for transition kernels $\mathbf{T}$, $\mathbf{T}_{I}$ and $\widehat{\mathbf{T}}_{I}$ respectively by showing an uniform minorization condition hold \citep{atchade2014bayesian}. Under the assumptions that $\bm{\Theta}$ is compact and $\mathcal{X}$ is finite, we can find a probability density $\lambda_{q}$, where $\epsilon_q = \inf_{\bm{\theta},\bm{\theta_{\ast}}} \frac{q(\bm{\theta},\bm{\theta_{\ast}})}{\lambda_{q}(\bm{\theta_{\ast}})} >0$. And there exists $C_{\alpha}=\inf_{\bm{\theta},\bm{\theta_{\ast}}} \alpha(\bm{\theta},\bm{\theta_{\ast}}) > 0$. Therefore, 
\begin{equation}
\mathbf{T}(\bm{\theta},A) \geq C_{\alpha}\epsilon_{q}\int_{A}\lambda_{q}(d\bm{\theta_{\ast}})\int_{\mathcal{X}}h(dS_\mathbf{y}|\bm{\theta_{\ast}}).
\end{equation}
\noindent According to Theorem 16.0.2 in \cite{meyn1993markov}, $\mathbf{T}$ converges to the $\pi(\cdot)$ at a geometric rate as 
\begin{equation}
\|\mathbf{T}^{n}(\bm{\theta},\cdot)-\pi(\cdot) \|_{TV} \leq (1-C_{\alpha}\epsilon_{q})^{n}. 
\label{invariant}
\end{equation}
\noindent The argument is the same for $\mathbf{T}_{I}$ and $\widehat{\mathbf{T}}_{I}$. 

From Assumptions 1-2 in Theorem~1 we can derive bound of difference between $\mathbf{T}(\bm{\theta},\cdot)$ and $\mathbf{T}_{I}(\bm{\theta},\cdot)$ as 
\begin{equation}
\begin{split}
& \|\mathbf{T}(\bm{\theta},\cdot)-\mathbf{T}_{I}(\bm{\theta,\cdot}) \|_{TV}\\
& \leq \sup_{|v|\leq 1} \int_{\bm{\Theta}}q(\bm{\theta},d\bm{\theta_{\ast}}) \Big| \int_{\mathcal{X}}[h(dS_\mathbf{y}|\bm{\theta_{\ast}})-g(dS_\mathbf{y}|\bm{\theta_{\ast}})]\alpha(\bm{\theta},\bm{\theta_{\ast}}) v(\bm{\theta},S_\mathbf{y})\Big|\\
& \leq S_{\alpha} \max(|K_{h}-1/c_{g}|,|k_{h}-c_{g}|) \int_{\bm{\Theta}} q(\bm{\theta},d\bm{\theta_{\ast}}),
\end{split}
\label{approx1}
\end{equation}
\noindent where $S_{\alpha}=\sup_{\bm{\theta},\bm{\theta_{\ast}} \in \bm{\Theta}} \sup_{S_{\mathbf{y}} \in \mathcal{X}} \alpha(\bm{\theta},\bm{\theta_{\ast}})$. Therefore $\|\mathbf{T}(\bm{\theta},\cdot)-\mathbf{T}_{I}(\bm{\theta,\cdot}) \|_{TV} $ is bounded for some constant $U < \infty $. 

Lastly we need to derive bound of difference between $\mathbf{T}_{I}(\bm{\theta},\cdot)$ and $\widehat{\mathbf{T}}_{I}(\bm{\theta,\cdot})$. Let $\nu(\bm{\theta})=\bm{\mu}(\bm{\theta})-\widehat{\bm{\mu}}(\bm{\theta})$ and let $\Pi(\bm{\theta})$ be a $p\times (p-1)$ matrix whose columns form a basis for the subspace orthogonal to $\nu(\bm{\theta})$. Let $\|A\|_{F}=\sqrt{\mbox{tr}(AA^\intercal)}$ be the Frobenius norm of the matrix $A$. According to Theorem 1.2 in \cite{devroye2018total}, total variation distance between two multivariate normal distribution is 
\begin{equation}
\|g(\cdot|\bm{\theta})-\widehat{g}(\cdot|\bm{\theta}) \|_{TV} \leq \frac{9}{2} \min \Big( 1, B_{M,d}(\bm{\theta}) \Big), 
\end{equation} 
where 
\begin{equation}
\begin{split}
& B_{M,d}(\bm{\theta}) = \max  \Big( \frac{ \nu(\bm{\theta})^\intercal (\bm{\Sigma}(\bm{\theta})-\widehat{\bm{\Sigma}}(\bm{\theta}))\nu(\bm{\theta}) }{ \nu(\bm{\theta})^\intercal \bm{\Sigma}(\bm{\theta})\nu(\bm{\theta}) },\\
& \frac{\nu(\bm{\theta})^\intercal \nu(\bm{\theta})}{\sqrt{\nu(\bm{\theta})^\intercal \bm{\Sigma}(\bm{\theta})\nu(\bm{\theta})}}\|(\Pi(\bm{\theta})\Sigma(\bm{\theta})\Pi(\bm{\theta}))^{-1}\Pi(\bm{\theta})\widehat{\Sigma}(\bm{\theta})\Pi(\bm{\theta})-I_{p-1}\|_{F}  \Big).
\end{split}
\end{equation}
\noindent Uisng this result, we can derive the bound of difference between $\mathbf{T}_{I}(\bm{\theta},\cdot)$ and $\widehat{\mathbf{T}}_{I}(\bm{\theta,\cdot})$ as
\begin{equation}
\begin{split}
& \|\mathbf{T}_{I}(\bm{\theta},\cdot)-\widehat{\mathbf{T}}_{I}(\bm{\theta,\cdot}) \|_{TV}\\
& \leq \sup_{|v|\leq 1} \int_{\bm{\Theta}}q(\bm{\theta},d\bm{\theta_{\ast}}) \Big| \int_{\mathcal{X}}[g(dS_\mathbf{y}|\bm{\theta_{\ast}})-\widehat{g}(dS_\mathbf{y}|\bm{\theta_{\ast}})]\alpha(\bm{\theta},\bm{\theta_{\ast}}) v(\bm{\theta},S_\mathbf{y})\Big|\\
& \leq \frac{9}{2}S_{\alpha} \int_{\bm{\Theta}} q(\bm{\theta},d\bm{\theta_{\ast}})\min \Big( 1, B_{M,d}(\bm{\theta}) \Big),
\end{split}
\label{approx2}
\end{equation}
\noindent where $S_{\alpha}=\sup_{\bm{\theta},\bm{\theta_{\ast}} \in \bm{\Theta}} \sup_{S_{\mathbf{y}} \in \mathcal{X}} \alpha(\bm{\theta},\bm{\theta_{\ast}})$. We note that $B_{M,d}(\bm{\theta})$ converges to 0 with increasing $M$ and $d$. This is because, we assume the parameter space $\bm{\Theta}$ is compact and $\widehat{\mu}(\bm{\theta})$, $\widehat{\Sigma}(\bm{\theta})$ converge to $\mu(\bm{\theta})$, $\Sigma(\bm{\theta})$ with increasing $M$. 

From \eqref{invariant}, \eqref{approx1} and \eqref{approx2} the approximation error for indirect auxiliary variable MCMC is
\begin{equation}
\begin{split}
\| \pi(\cdot) - \widehat{\mathbf{T}}^{n}_{I}(\bm{\theta},\cdot)  \|_{TV}  & \leq   \| \pi(\cdot) - \mathbf{T}^{n}(\bm{\theta},\cdot) \|_{TV}  + \| \mathbf{T}^{n}(\bm{\theta},\cdot) - \mathbf{T}^{n}_{I}(\bm{\theta},\cdot)  \|_{TV}\\
& + \| \mathbf{T}^{n}_{I}(\bm{\theta},\cdot) - \widehat{\mathbf{T}}^{n}_{I}(\bm{\theta},\cdot) \|_{TV} \\
& \leq (1-C_{\alpha}\epsilon_{q})^{n} + U + \epsilon(M) + \epsilon(d) 
\end{split}
\end{equation}
 
We note that Assumption 3 can be relaxed for point process applications. For point process models, the summary statistics of an auxiliary variable can be generated from the probability model $h(S_{\mathbf{y}}|\bm{\theta_{\ast}})$ via birth-death MCMC \citep{geyer1994simulation}. Consider the acceptance probability of the birth-death MCMC $\alpha(\mathbf{y},\mathbf{y}_{\ast})$. Once  $\inf_{\mathbf{y},\mathbf{y}_{\ast}\in \mathcal{X}}\alpha(\mathbf{y},\mathbf{y}_{\ast}) > 0$ is satisfied, we can show Theorem~1 similarly. This holds for irreducible birth-death MCMC, which is available in many applications.

\bibliography{Reference}
\end{document}